\documentclass[10pt,prl,aps,twocolumn,showpacs,nofootinbib,superscriptaddress,floatfix,nopreprintnumbers]{revtex4-1}

\usepackage{amsmath,graphicx,mathrsfs,amsfonts}
\usepackage{bm} 
\usepackage{subfigure}
\usepackage[colorlinks=true,linktocpage=true,linkcolor=blue,citecolor=blue]{hyperref}
\usepackage[usenames,dvipsnames]{color}

\begin{document}

\title{
Hydrodynamics from quantum fields: a regularized expansion from the Wigner distribution.
}

\author{L. Tinti}
\affiliation{Institut f\"ur Theoretische Physik, Johann Wolfgang Goethe-Universit\"at, Max-von-Laue-Str.~1, D-60438 Frankfurt am Main, Germany}

\begin{abstract}

Second-order relativistic hydrodynamics is surprisingly predictive, even in the presence of large gradients. The hydrodynamic expansion from the method of moments does not require a gradient expansion, but it is intrinsically bound to the classic nature of relativistic kinetic theory. In this work a modified version of the method of moments is applied the Wigner distribution (the quantum precursor of the distribution function) to recover a systematically improvable hydrodynamic expansion, avoiding the divergences that would otherwise appear in the quantum case. The convergence of the regularized expansion is checked numerically in a far from equilibrium, distant from the kinetic limit case.

\end{abstract}

\pacs{12.38.Mh, 24.10.Nz, 25.75.-q, 47.10.+g}

\keywords{relativistic heavy-ion collisions, viscous hydrodynamics, Wigner distribution, hydrodynamic expansion}

\maketitle


\noindent \textsl{1. Introduction.}  Relativistic hydrodynamics has been used in a wide range of physical systems, from astrophysical plasmas to heavy-ion collisions \cite{Bertschinger:1998tv,Mizuno:2015yxa,Heinz:2013th}. It is often thought to be inescapably an expansion in gradients,  or an approximation of the relativistic Boltzmann equation. Both instances are not general enough to explain the hydrodynamic behavior, namely, of the quark-gluon plasma in the cross-over region. The Chapman-Enskog expansion \cite{ChapmanCowling} involves a systematic power counting of the gradients of the hydrodynamic variables. It can be performed in the context of relativistic kinetic theory~\cite{kremer} (weak coupling) as well as strongly coupled relativistic systems~\cite{Bhattacharyya:2008jc,Baier:2007ix}. The leading order corresponds to ideal hydrodynamics, and the first order to the relativistic Navier-Stokes equations \cite{LandauLifshitzFluids}. The latter equations, in general, violate causality \cite{PichonViscous} and are unstable~\cite{Hiscock_Lindblom_stability_1983,Hiscock_Lindblom_instability_1985,Denicol:2008ha,Pu:2009fj}, but it has been proven~\cite{Bemfica:2017wps} that it is possible to obtain causal and stable theory {\it at first order in gradients} if different definitions for the hydrodynamic fields are used. The most common way to preserve cusality and stability, however, is to add second order corrections~\cite{Bhattacharyya:2008jc,Baier:2007ix}. Third-order terms can be included \cite{GrozdanovKaplisThirdOrder, Jaiswal:2013vta}, though precise statements regarding causality and stability are not available at higher orders. Recent works pointed out that the gradient series has zero radius of convergence~\cite{Heller:2013fn,Buchel:2016cbj,Denicol:2016bjh,Heller:2016rtz}. The lowest orders of an asymptotic series can be numerically accurate, as it seems the case for second-order relativistic hydrodynamics. However, in the absence of a fast convergence, one cannot look at the next orders to guess the accuracy of the approximation. Second-order relativistic hydrodynamics can be obtained independently as an approximation of relativistic kinetic theory~\cite{Denicol:2012cn}. The method of moments can be used to systematically improve the hydrodynamic expansion to higher orders. Under flow conditions of extreme symmetry, in which the relativistic Boltzmann equation can be solved exactly \cite{Baym:1984np, Florkowski:2013lza, Florkowski:2013lya, Denicol:2014xca, Denicol:2014tha}, this procedure has been shown to converge rapidly to the exact results of relativistic kinetic theory \cite{Denicol:2014mca, Denicol:2016bjh, Bazow:2016oky}. The relativistic Boltzmann equation itself, however, is expected to be a valid approximation only for weakly interacting asymptotic particle states~\cite{degroot}. Close to the cross-over region it is not clear what are (if they exist at all) the correct particle-like degrees of freedom.

On the other hand, the Wigner distribution is well defined in quantum field theory, it fulfills an off-shell version of the relativistic Boltzmann equation, and it provides the distribution function in the kinetic limit.~\cite{degroot}. An often invoked, and yet not properly formalized, justification of hydrodynamics relies on the similarities between the two: from the same structure of the fundamental equations, one should be able to extract similar results. The purpose of this work is to show that, after a proper regularization of the integrals, one can generalize the method of moments and recover a systematically improvable hydrodynamic expansion.



\noindent \textsl{2.Moment expansion.} It is possible to extract the exact comoving derivative of the distribution function $u\cdot \partial f=\dot f$, directly from the relativistic Boltzmann equation

\begin{equation}\label{Boltzmann}
 \begin{split}
 p\cdot\partial f(x,p) &= - {\cal C}[f] \Rightarrow (p\cdot u) u\cdot f = -p\cdot\nabla f -{\cal C} \\
\dot f &=-\frac{1}{(p\cdot u)}\left( \vphantom{\frac{}{}}p\cdot \nabla f + {\cal C} \right),
 \end{split}
\end{equation}
indepedently of the particular definition of the four-velocity $u^\mu$ and the collisional kernel ${\cal C}$ . The gradient orthogonal to $u^\mu$ being

\begin{equation}
 \nabla_\mu =\Delta^\nu_\mu\partial_\nu = \left(g^{\mu\nu}-u^\mu u^\nu\right)\partial_\mu.
\end{equation}
The {\it mostly minus} convention for the metric is in use $g={\rm diag}(1,-1,-1,-1)$. The tensor moments of the distribution function read

\begin{equation}\label{reducible}
 \begin{split}
  {\cal F}_r^{\mu_1\cdots\mu_s} &= \int_{\bf p} (p\cdot u)^r p^{\mu_1}\cdots p^{\mu_s} f,
 \end{split}
\end{equation}
with $\int_{\bf p}$ the covariant (on-shell) momentum integral

\begin{equation}
 \begin{split}
  \int_{\bf p} &= \!\frac{N_{\rm dof}}{(2\pi)^3}\int \!\frac{d^3 p}{ E_{\bf p}} = \! \frac{N_{\rm dof}}{(2\pi)^3}\int \!\!d^4p \,  2\theta(E)\delta(p^2-m^2),
 \end{split}
\end{equation}
and $N_{\rm dof}$ the eventual degeneracy factor.

Making use of~(\ref{Boltzmann}), after some straightforward algebra, one can find the exact comoving derivatives of the moments~(\ref{reducible})

\begin{equation}\label{red_ev_cl}
 \begin{split}
  \dot{\cal F}^{\mu_1\cdots \mu_s}_r +C_{r-1}^{\mu_1\cdots\mu_s} &= r \dot u_\alpha {\cal F}_{r-1}^{\alpha\mu_1\cdots\mu_s} -\nabla_\alpha {\cal F}_{r-1}^{\alpha\mu_1\cdots\mu_s } \\
  & \qquad  +(r-1)\nabla_\alpha u_\beta {\cal F}^{\alpha \beta \mu_1\cdots \mu_s}_{r-2},
  \end{split}
\end{equation}
in which $C_{r-1}^{\mu_1\cdots\mu_s}$ is a short-hand notation for

\begin{equation}\label{coll_on_shell}
 \begin{split}
  {C}_{r-1}^{\mu_1\cdots\mu_s} &= \int_{\bf p} (p\cdot u)^{r-1} p^{\mu_1}\cdots p^{\mu_s} {\cal C}[f].
 \end{split}
\end{equation}
In particular, the exact evolution of the stress-energy tensor  $T^{\mu\nu}={\cal F}^{\mu\nu}_0$ reads

\begin{equation}\label{Tmunu_Boltzmann}
 \begin{split}
  &\dot T^{\mu\nu} + C_{-1}^{\mu\nu} = -\nabla_\alpha {\cal F}_{-1}^{\alpha\mu\nu} - \nabla_\alpha u_\beta {\cal F}^{\alpha\beta\mu\nu}_{-2}.
 \end{split}
\end{equation}
%
%
%
%
%
The contraction of the last equation with the four-velocity $u_\mu$ is the local conservation of four-momentum $\partial_\mu T^{\mu\nu}=0$, which is included at all orders of the hydrodynamic  expansion. The remaining equations provide the exact evolution of the pressure tensor

\begin{equation}\label{P_ev_Bol}
 \begin{split}
 &  \dot {\cal P}^{\langle\mu\rangle\langle\nu\rangle}+ C_1^{\langle\mu\rangle\langle\nu\rangle} = -{\cal P}^{\mu\alpha}\nabla_\alpha u^\nu -{\cal P}^{\nu\alpha}\nabla_\alpha u^\mu \\
  &  - \theta {\cal P}^{\mu\nu} +  q^\mu \dot u^\mu + \dot u^\mu q^\nu-\nabla_\alpha {\mathfrak f}_{-1}^{\alpha\langle\mu\rangle\langle\nu\rangle} - \nabla_\alpha u_\beta {\mathfrak f}^{\alpha\beta\mu\nu}_{-2}.
 \end{split}
\end{equation}
Both instances can be obtained without approximations from Eq.~(\ref{Tmunu_Boltzmann}) making use of the exact relations

\begin{equation}\label{exact_reducible}
 \begin{split}
 u_{\nu_1}\cdots u_{\nu_n} {\cal F}^{\nu_1\cdots\nu_n\mu_1\cdots\mu_s}_r &= {\cal F}^{\mu_1\cdots\mu_s}_{r+n}, 
 \end{split}
\end{equation}
and the similar ones for the collisional integrals. The brackets in equation~(\ref{P_ev_Bol}) represent the projection ${\cal O}^{\langle\mu\rangle\cdots} =\Delta^\mu_\nu {\cal O}^{\nu\cdots}$, and the shorthand notation

\begin{equation}
 {\mathfrak f}_r^{\mu_1\cdots\mu_s}={\cal F}_r^{\langle\mu_1\rangle\cdots\langle\mu_s\rangle},
\end{equation}
is in use. The energy flux is then $q^\mu = {\mathfrak f}_1^\mu$ (vanishing in the Landau frame), and ${\mathfrak f}_0^{\mu\nu}$ is the pressure tensor, often decomposed in the following way

\begin{equation}
 {\mathfrak f}_0^{\mu\nu} = {\cal P}^{\mu\nu} = -\left({\cal P} +\Pi\right)\Delta^{\mu\nu} +\pi^{\mu\nu},
\end{equation}
being ${\cal P}$ the hydrostatic pressure, $\Pi$ the bulk pressure correction and $\pi^{\mu\nu}$ the shear one (space-like and trace-less). On the right hand side of Eq.~(\ref{P_ev_Bol}) all terms except the last two are components of $T^{\mu\nu}$. One can either approximate these tensors (second-order relativistic hydrodynamics or modified versions of it~\cite{Denicol:2014loa,Tinti:2015xwa,Molnar:2016gwq}) or treat them as independent degrees of freedom and use Eq.~(\ref{red_ev_cl}) for their evolution~\cite{Denicol:2014loa,Tinti:2015xwa,Molnar:2016gwq}. The same arguments hold for the collisional integral in the left hand side of Eq.~(\ref{P_ev_Bol}), and these steps can be repeated further to get the higher orders of the expansion.

\noindent \textsl{3.The quantum case.} In its most simple instance, the Wigner distribution is the Fourier transform of the two point expectation value~\cite{degroot}

\begin{equation}\label{Wigner}
 W(x,k)= \int \frac{d^4 v}{(2\pi)^4} e^{-i k\cdot v} \langle \phi^\dagger(x+\frac{1}{2}v) \phi(x -\frac{1}{2}v)\rangle,
\end{equation}
with the expectation value for a generic state (pure or mixed) of the system. The stress-energy tensor reads

 \begin{equation}\label{T_scalar}
  T^{\mu\nu} = \int \!\! d^4 k\; k^\mu k^\nu W(x,k).
 \end{equation}
The Wigner distribution satisfies an equation very similar to Eq.~(\ref{Boltzmann}), which can be derived from the field equations

\begin{equation}\label{Q_kin}
 k\cdot \partial W= -{\cal C}_W[W,\cdots] \Rightarrow (k\cdot u) \dot W = -k\cdot\nabla W -{\cal C}_W.
\end{equation}
The quantum version of the collisional kernel ${\cal C}_W$ depends on the interaction, and it vanishes for free fields.
 Because of the similarities with relativistic kinematics, one would expect to recover the hydrodynamics expansion with the minimal substitution

\begin{equation}
 \begin{split}
  \int_{\bf p}\to \int d^4 k , \qquad \frac{N_{\rm dof}}{(2\pi)^3}f(x,p)\to W(x,k).
 \end{split}
\end{equation}
%
Following the steps outlined in the last section, one would recover Eq.~(\ref{P_ev_Bol}), however the last two integrals are ill-defined in the quantum case, for instance
\begin{equation}\label{f_-1_W}
  {\mathfrak f}_{-1}^{\alpha\mu\nu} = \int d^4 k \frac{k^{\langle\alpha\rangle} k^{\langle\mu\rangle} k^{\langle\nu\rangle}}{k\cdot u} W(x,k),
\end{equation}
the Wigner distribution is not on shell in general, and equation~(\ref{f_-1_W}) diverges even if $W(x,k)$ is proportional to an arbitrarily sharp Gaussian around the on-shell energy, instead of an actual delta distribution.




It is possible to revisit a regularization scheme introduced in~\cite{Tinti:2018qfb,Tinti:2018nrp} to deal with similar infrared divergences, appearing at higher orders of the expansion for the Boltzmann-Vlasov equation. The main objects are the regularized tensors

\begin{equation}\label{phi}
 \phi^{\mu_1\cdots\mu_s}_n =\!\int \!\! d^4 k \; (k\cdot u)^n e^{-\zeta(k\cdot u)^2} k^{\langle\mu_1\rangle}\cdots k^{\langle\mu_s\rangle}W,
\end{equation}
with $\zeta\ge 0$ a parameter with the dimensions of a length squared. One recovers all  the well defined ${\mathfrak f}_{n-2}^{\mu_1\cdots \mu_s}$ moments, after integration integrating appropriate regularized tensor from $\zeta=0$ to infinity, including the components of $T^{\mu\nu}$. From~(\ref{Q_kin}) one obtains the dynamical equations

\begin{widetext}
\begin{equation}\label{Q_ev_n}
 \begin{split}
  &\dot\phi^{\langle\mu_1\rangle\cdots \langle\mu_s\rangle}_n +{\tilde C_{n-1}}^{\langle\mu_1\rangle\cdots\langle\mu_s\rangle} = s \dot u^{(\mu_1}\phi_{n+1}^{\mu_2\cdots\mu_s)} -\theta \phi^{\mu_1\cdots \mu_s}_n -s \nabla_\alpha u^{(\mu_1}\phi_n^{\mu_2\cdots\mu_s) \alpha}\\
 & -\nabla_\alpha \phi_{n-1}^{\alpha\langle\mu_1\rangle\cdots\langle\mu_s \rangle} + \dot u_\alpha\left[n \phi_{n-1}^{\alpha\mu_1\cdots\mu_s} -2\zeta\phi_{n+1}^{\alpha\mu_1\cdots\mu_s}\right] +\nabla_\alpha u_\beta\left[ (n-1)\phi^{\alpha \beta \mu_1\cdots \mu_s}_{n-2} -2\zeta\phi^{\alpha \beta \mu_1\cdots \mu_s}_{n} \right],
 \end{split}
\end{equation}
\end{widetext}
being $\theta = \nabla_\mu u^\mu$ the scalar expansion, and ${\tilde C_{n-1}}^{\langle\mu_1\rangle\cdots\langle\mu_s\rangle}$ having the same prescription of the regularized integrals in~(\ref{phi}) weighted with ${\cal C}_W$ instead of $W$.
%
%
%
For $n\ge 1$ no diverging integral appears on the left hand side of Eq.~(\ref{Q_ev_n}), and one can check that in the kinetic limit (after integrating over $\zeta$) one recovers all the the classical equations in~(\ref{red_ev_cl}). 
The exact evolution of the pressure tensor ${\cal P}^{\mu\nu}=\int d\zeta \; \phi_2^{\mu\nu}$ then reads

\begin{widetext}
\begin{equation}\label{P_ev}
 \begin{split}
  \dot {\cal P}^{\langle\mu\rangle\langle\nu\rangle} &= -{\cal P}^{\mu\alpha}\nabla_\alpha u^\nu -{\cal P}^{\nu\alpha}\nabla_\alpha u^\mu  +  q^\mu \dot u^\mu + \dot u^\mu q^\nu \\
  & \quad +\int_0^\infty\!\!\!d\zeta\left\{-{\tilde C}_1^{\langle\mu\rangle\langle\nu\rangle} -\nabla_\alpha \phi_1^{\alpha\langle\mu\rangle\langle\nu\rangle} +\nabla_\alpha u_\beta \left[ \phi^{\alpha\beta\mu\nu}_0 -2\zeta \phi_2^{\alpha\beta\mu\nu}\right]\right\}.
 \end{split}
\end{equation}
\end{widetext}
This equation is well defined, however one must be careful and avoid to to split the last integral. It has to be convergent, but it but it doesn't have to converge \textit{uniformly}. It can still be approximated to obtain second-order viscous hydrodynamics (e.g. substituting the higher ranking tensors with their equilibrium expectation values). Otherwise one can treat the new degrees of freedom as dynamical variables, using Eq.~(\ref{Q_ev_n}) for their evolution, and so on for the higher orders. Since, from the very definition~(\ref{phi}), the following relation holds

\begin{equation}
 \int_\zeta^\infty d\zeta^\prime \phi_{n+2}^{\mu_1\cdots\mu_s}(x,\zeta^\prime)=\phi_{n}^{\mu_1\cdots\mu_s}(x,\zeta),
\end{equation}
one can use a limited number of generations $n$, as it has been done in the classical case~\cite{Tinti:2018qfb,Tinti:2018nrp}. More interestingly, if the situation is tame enough, it is possible to perform exactly the integration in the right hand side and obtain a set of well defined,  $\zeta$-independent sources for the evolution of the pressure tensor, and for the higher order equations. In fact, this happens in the system used in this work for the numerical comparisons. Equations~(\ref{Wigner}) and~(\ref{T_scalar}) can be more complicated in general (see, for instance~\cite{Elze:1986hq}). It is important to note, however, that the very same infrared divergences appear, and they can be handled with the same regularization scheme.

\noindent \textsl{3.Comparison with the exact results.} 
It is possible to generalize the approach in~\cite{Florkowski:2013lya} to the quantum case, obtaining an exact solution of the Wigner distribution in a very symmetric case. The generalized collisional kernel ${\cal C}_W$ is treated in the relaxation time approximation, and the system is assumed to be invariant under a generalized Bjorken symmetry

\begin{equation}\label{Simplification}
 \begin{split}
  & W(x, k) = W(\tau, k_T, {w}^2,{v}^2) \\
  & k\cdot \partial W = -\frac{(k\cdot u)}{\tau_R}\left[ \vphantom{\frac{}{}}W-W_{\rm eq.} \right] \equiv -\frac{(k\cdot u)}{\tau_R}\delta W.
 \end{split}
\end{equation}
The same notation is in use as in~\cite{Florkowski:2013lya}

\begin{equation}\label{variables}
 \begin{split}
  & \tau =\sqrt{t^2-z^2},\quad k_T =\sqrt{(k^x)^2+(k^y)^2}, \\
  &  v = tk^0 - z k^z, \quad w= zk^0-t k^z.
 \end{split}
\end{equation}
The Wigner distribution, differently from the classical counterpart, has an explicit dependence on $v$, since $k^\mu$ is not on-shell in general, and the frequency $k^0$ (hence $v$ itself) is not dictated by the other variables. The explicit dependence on $w^2$ is to make manifest the invariance for axis reflections, while the dependence on $v^2$ rather than $v$ does not come from the Bjorken symmery, and it is a further mathematical simplifications. Among the other things, it entails that the charge density in the comoving frame is vanishing (no net charge). Considering that the only time-like four vector consistent with the symmetry is $u^\mu =(t,0,0,z)/\tau$ and using the definitions~(\ref{variables}), the evolution~(\ref{Simplification}) simplifies

\begin{equation}\label{Simplified}
 \left[\partial_\tau + 2\frac{v^2-w^2}{\tau}\partial_{v^2}\right] W = -\frac{1}{\tau_R}\delta W.
\end{equation}
%
%

Just like in the classical case, the conservation of energy an momentum requires the effective temperature be the one from the the Landau matching. Choosing for the equilibrium distribution and relaxation time

\begin{equation}\label{choice}
 W_{\rm eq.} = \frac{2\delta(k^2)}{(2\pi)^3} e^{-\frac{1}{T(\tau)}\sqrt{k_T^2 +\frac{w^2}{\tau^2}}}, \qquad \tau_R = \frac{5\bar \eta}{T(\tau)},
\end{equation}
one has
\begin{equation}
 {\cal E}(\tau) = {\cal E}_{\rm eq.}(T) = \frac{6}{\pi^2}T^4\Rightarrow T(\tau) =\left(\frac{\pi^2 {\cal E}(\tau)}{6}\right)^{\frac{1}{4}}.
\end{equation}
This choice~(\ref{choice}) is mainly for mathematical simplicity. It entails that at equilibrium the system has a conformal equation of state, and in the kinetic limit it has a constant ratio of shear viscosity over entropy $\bar \eta$. Using the method of characteristics it is possible to write an implicit solution of~(\ref{Simplified})

\begin{equation}\label{W_sol}
\!\!\! W \!\!= \!D(\tau,\tau_0)W_{\rm f.s.} \!+\! \frac{2\delta(k^2)}{(2\pi)^3}\int_{\tau_0}^\tau \!\!\!ds \frac{D(\tau,s)}{\tau_R(s)}e^{-\frac{\sqrt{k_T^2 +\frac{w^2}{s^2}}}{T(s)}},
\end{equation}
being the damping function $D$ and the free-streaming $W_{\rm f.s.}$

\begin{equation}
 \begin{split}
  & D(\tau_1,\tau_2)=e^{-\int_{\tau_1}^{\tau_2}\frac{ds}{\tau_R(s)}}, \\
  & W_{\rm f.s.}(\tau,k_T, w^2, v^2) = W_0(k_T,w^2,v_0^2), \\
  & v_0^2 =v^2 \left(\frac{\tau_0}{\tau}\right)^2 +w^2\frac{\tau^2 -\tau_0^2}{\tau^2},
 \end{split}
\end{equation}
for a generic initial condition $W_0 =W(\tau=\tau_0)$. It is possible then to use the self consistency method used in Refs.~\cite{Banerjee:1989by,Florkowski:2013lya,Denicol:2014tha} to obtain the numerical values of~(\ref{W_sol}) up to an arbitrary precision. Because of the strong symmetry of the system, the non-trivial components of the stress-energy tensor are the proper energy density, the longitudinal pressure ${\cal P}_L = T^{\mu\nu}z_\mu z_\nu$, the transverse pressure ${\cal P}_T=T^{\mu\nu}x_\mu x_\nu$, with $x^\mu=(0,1,0,0)$ and $z^\mu=(z,0,0,t)/\tau$. The only moments~(\ref{phi}) related to them, directly or indirectly, are

\begin{equation}
 \begin{split}
  & L_n(\tau,\zeta) = \phi_2^{\mu_1\cdots\mu_n}z_{\mu_1}\cdots z_{\mu_n}, \\
  & T_n(\tau,\zeta) = \phi_2^{\alpha_1\alpha_2\mu_1\cdots\mu_{n}}x_{\alpha_1}x_{\alpha_2}z_{\mu_1}\cdots z_{\mu_n},
 \end{split}
\end{equation}
for instance, ${\cal E}(\tau)=L_0(\tau,\zeta=0)$, ${\cal P}_{T} =\int d\zeta T_0$, ${\cal P}_L = \int d\zeta L_1$. Plugging~(\ref{Simplification}) into~(\ref{Q_ev_n})

\begin{equation}\label{L_T_ev}
 \begin{split}
 & \dot L_n +\frac{1}{\tau_R}\delta L_n = -\frac{2n+1}{\tau} L_n +\frac{1}{\tau}\hat{\cal L}L_{n+1}, \\
 & \dot T_n +\frac{1}{\tau_R}\delta T_n = -\frac{2n+1}{\tau} L_n +\frac{1}{\tau}\hat{\cal L}T_{n+1},
 \end{split}
\end{equation}
being $\delta L_n$ and $\delta T_n$ the difference between the moments and their local equilibrium expectation value. It is possible to directly integrate in $\zeta$ the non-hydrodynamic sources on the right hand side of~(\ref{P_ev}), and obtain $\zeta$-independent set of equations.  Introducing the linear operator
\begin{equation}
 \hat{\cal L} f(\zeta)= 2\zeta f(\zeta) -\int_\zeta^\infty d\zeta^\prime f(\zeta^\prime),
\end{equation}
the dynamical equations for $T^{\mu\nu}$ then read

\begin{equation}\label{hydro}
 \begin{split}
  & \dot {\cal E} = -\frac{1}{\tau}\left( \vphantom{\frac{}{}}{\cal E} +{\cal P}_L \right), \\
  & \dot {\cal P}_L +\frac{1}{\tau_R}\delta {\cal P}_L = -\frac{3}{\tau} {\cal P}_L +\frac{1}{\tau}{\cal R}_L^{(1)}, \\
  & \dot {\cal P}_T +\frac{1}{\tau_R}\delta {\cal P}_T = -\frac{1}{\tau} {\cal P}_T +\frac{1}{\tau}{\cal R}_T^{(1)},
 \end{split}
\end{equation}
the residual moments (an their equilibrium expectation value) being

\begin{equation}
 \begin{split}
  {\cal R}_L^{(n)} &= \int_0^\infty \!\!\!\! d\zeta \left[( \hat{\cal L})^n L_{n+1}\right] \stackrel{{\rm eq.}}{\longrightarrow} \frac{(2n-1)!!}{2n+3}{\cal E},\\
  {\cal R}_T^{(n)} &= \int_0^\infty \!\!\!\! d\zeta \left[( \hat{\cal L})^n T_{n}\right] \stackrel{{\rm eq.}}{\longrightarrow} \frac{(2n-1)!!}{(2n+3)(2n+1)}{\cal E},
 \end{split}
\end{equation}
 and their evolution, stemming directly from~(\ref{L_T_ev})

\begin{equation}\label{extra}
 \begin{split}
  & \dot {\cal R}_L^{(n)} +\frac{1}{\tau_R}\delta {\cal R}_L^{(n)} = -\frac{2n+3}{\tau} {\cal R}_L^{(n)} +\frac{1}{\tau}{\cal R}_L^{(n+1)}, \\
  & \dot {\cal R}_T^{(n)} +\frac{1}{\tau_R}\delta {\cal R}_T^{(n)} = -\frac{2n+1}{\tau} {\cal R}_T^{(n)} +\frac{1}{\tau}{\cal R}_T^{(n+1)}.
 \end{split}
\end{equation}
Second-order viscous hydrodynamics corresponds to taking only~(\ref{hydro}) as the dynamical equations, and substituting the residual moments, e.g., with their equilibrium expectation values. For the higher orders one considers the residual moments up to a maximum $n$ as dynamical variables, evolving according to~(\ref{extra}) and approximating the $(n+1)$-residual moments.

An interesting initial condition is

\begin{equation}\label{initial}
 W_0 = \frac{2}{(2\pi)^3} \frac{e^{-\frac{v^2}{2\tau_0^2 \sigma}-\frac{\sqrt{\sigma}}{T_{\rm in}}}}{\sqrt{2\pi}\sigma}\left[ 1-3 P_2\left(\frac{w}{\sqrt{\sigma}}\right) \right],
\end{equation}
being $\sigma\equiv k_T^2 -\frac{w^2}{\tau_0^2}$ and $P_2(x)$ the second Laguerre polynomial. The initial values of the energy density and pressure then read

\begin{equation}\label{initial_hydro}
 \begin{split}
  {\cal E}(\tau_0)& = \frac{6}{\pi^2}T_{\rm in}^4, \\
  {\cal P}_L(\tau_0) &= -\frac{1}{15}{\cal E}(\tau_0), \quad
  {\cal P}_T(\tau_0) =\frac{8}{15}{\cal E}(\tau_0),
 \end{split}
\end{equation}
while for the non-hydrodynamic moments


%
\begin{equation}\label{initial_non_hydro}
 \begin{split}
  {\cal R}_L^{(n)}(\tau_0) &= -\frac{4n+1}{2n+5}\frac{(-1)^n}{2n+3}{\cal E}(\tau_0) , \\
  {\cal R}_T^{(n)}(\tau_0) &=-\frac{4n-8}{2n+5}\frac{(-1)^n}{(2n+3)(2n+1)}{\cal E}(\tau_0).
 \end{split}
\end{equation}
The distribution is very far from the kinetic limit and it remains off-shell during the whole evolution. Selecting an initial temperature $T_{\rm in}= 600 MeV$, $4\pi \bar \eta=3$ and $\tau_0 =0.25 fm/c$, one has similar conditions to the ones in heavy-ion collisions in which hydrodynamics is applied. In the framework of a gradient expansion, one would expect hydrodynamics to fail, because of the large pressure corrections and large gradients ($\tau_R(\tau_0)\theta(\tau_0)\simeq 1.6$). On the other hand, the regularized expansion presented in this work shows a different picture. One can estimate the relative error committed approximating the residual moments in the right hand side of Eq.~(\ref{hydro}) with their equilibrium expectation values. At the beginning of the expansion there is a sizable error $\Delta{\cal R}^{(1)}_L/{\cal R}^{(1)}_L = 40\%$ and a very large $\Delta{\cal R}^{(1)}_T/{\cal R}^{(1)}_T = 275\%$. However, the self-coupling of the hydrodynamic degrees of freedom is dominant, and there is a moderate $\Delta \dot{\cal P}_L/\dot{\cal P}_L \simeq 22\%$ and $\Delta\dot{\cal P}/\dot{\cal P}_T \simeq 15 \%$. It is reasonable, then to expect a qualitative agreement with the exact results. It is interesting to note, though, that in a composite quantity like the trace anomaly $T^{\mu\nu}g_{\mu\nu}$, the self coupling part of the hydrodynamic quantities compensate and one has $\simeq121\%$ error in its time derivative at the beginning. Making it likely that, differently from the energy density and the anisotropy ${\cal P}_L/{\cal P}_T$, the trace anomaly won't be well reproduced by hydrodynamics.

%
\begin{figure}
\includegraphics[angle=0,width=\columnwidth]{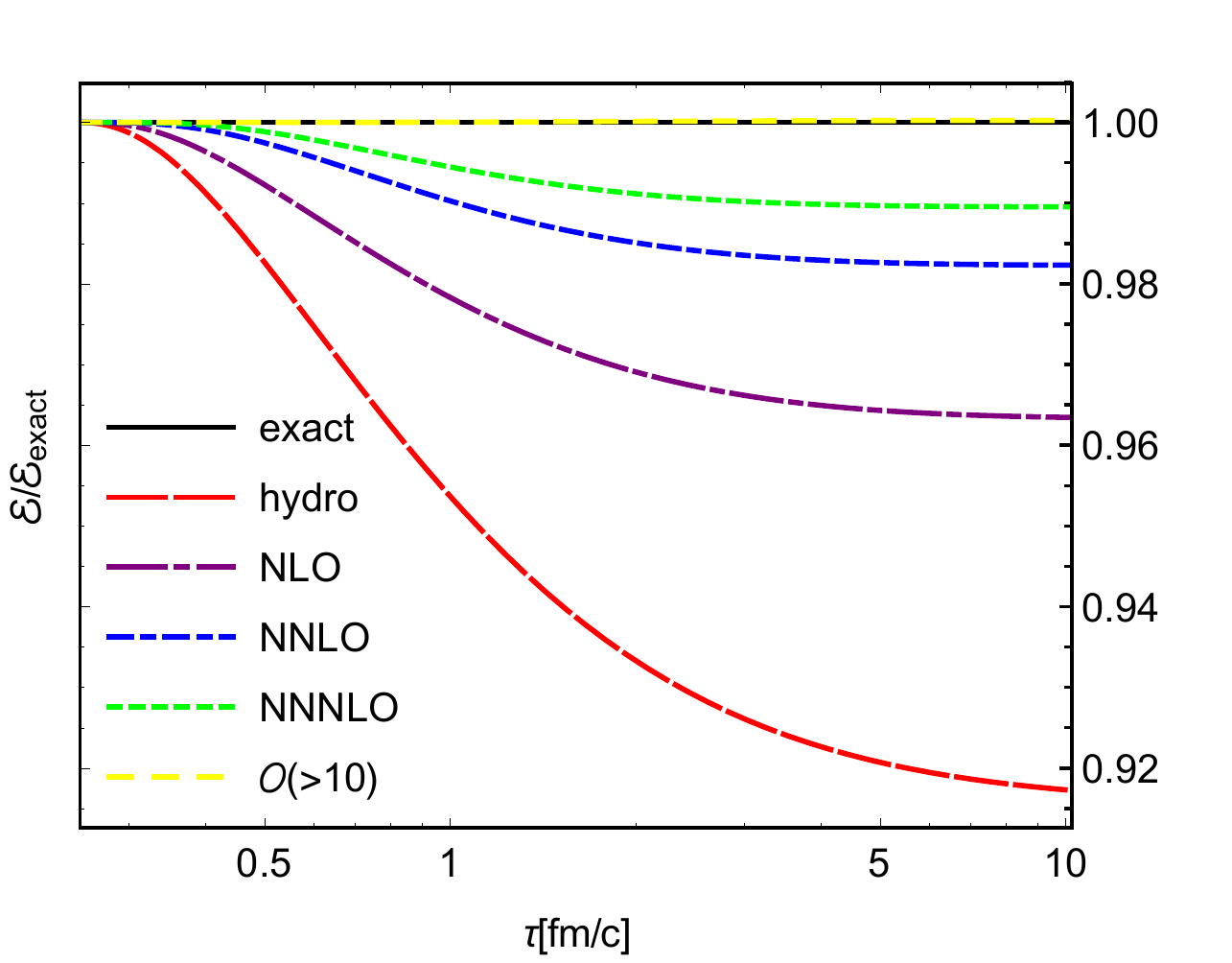}

\caption{(Color online) Ratio of the energy density over the exact energy density. The exact results in the black solid line, hydrodynamics (red dashed line), and the next orders: NLO (purple dash-dotted), NNLO (blue dash-dotted)and  NNLO (green dotted), $10$'th order or higher (yellow dotted). 
}
\label{F1}
\end{figure}
\begin{figure}
\includegraphics[angle=0,width=\columnwidth]{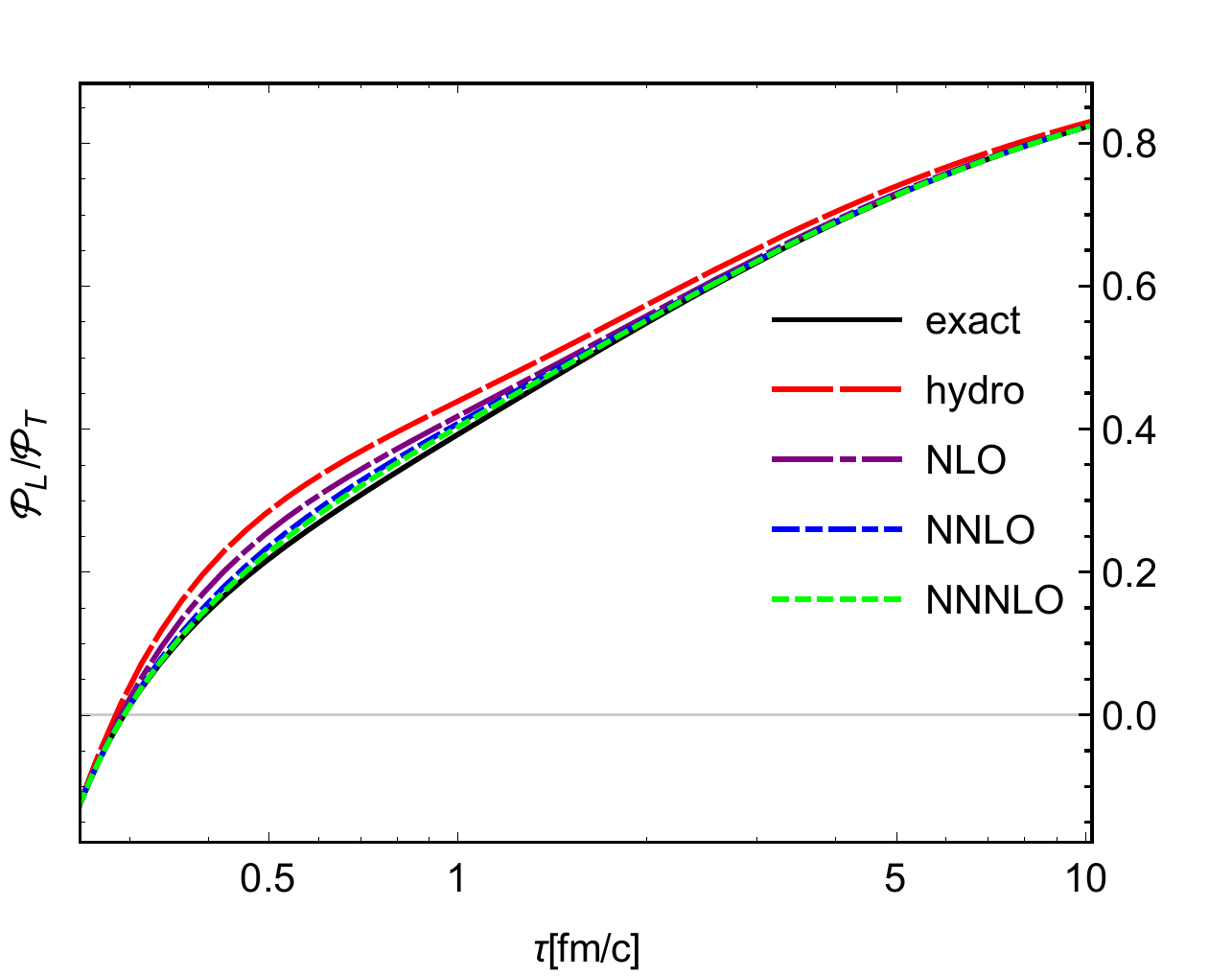}

\caption{
	(Color online) Anisotropy evolution. The exact results in the black solid line, hydrodynamics (red dashed line), and the next orders. NLO (purple dash-dotted), NNLO (blue dash-dotted)and  NNLO (green dotted). 
}
\label{F2}
\end{figure}
Indeed, as shown in Fig.~\ref{F1} and Fig.~\ref{F2}, the energy density and the anisotropy are relatively well reproduced; while the trace anomaly, Fig.~\ref{F3}, is largely overestimated. A definitive answer on the convergence of the regularized expansion is hard to get from a theoretical point of view. However the system is simple enough to numerically check the higher orders. In Figs.~{\ref{F1}, \ref{F2}, and \ref{F3} it is shown that each additional step improves the accuracy. At the $10$'th order and higher there is  substantially no difference with the exact solutions (I have been checking up to the $100$'th order). 

\begin{figure}
\includegraphics[angle=0,width=\columnwidth]{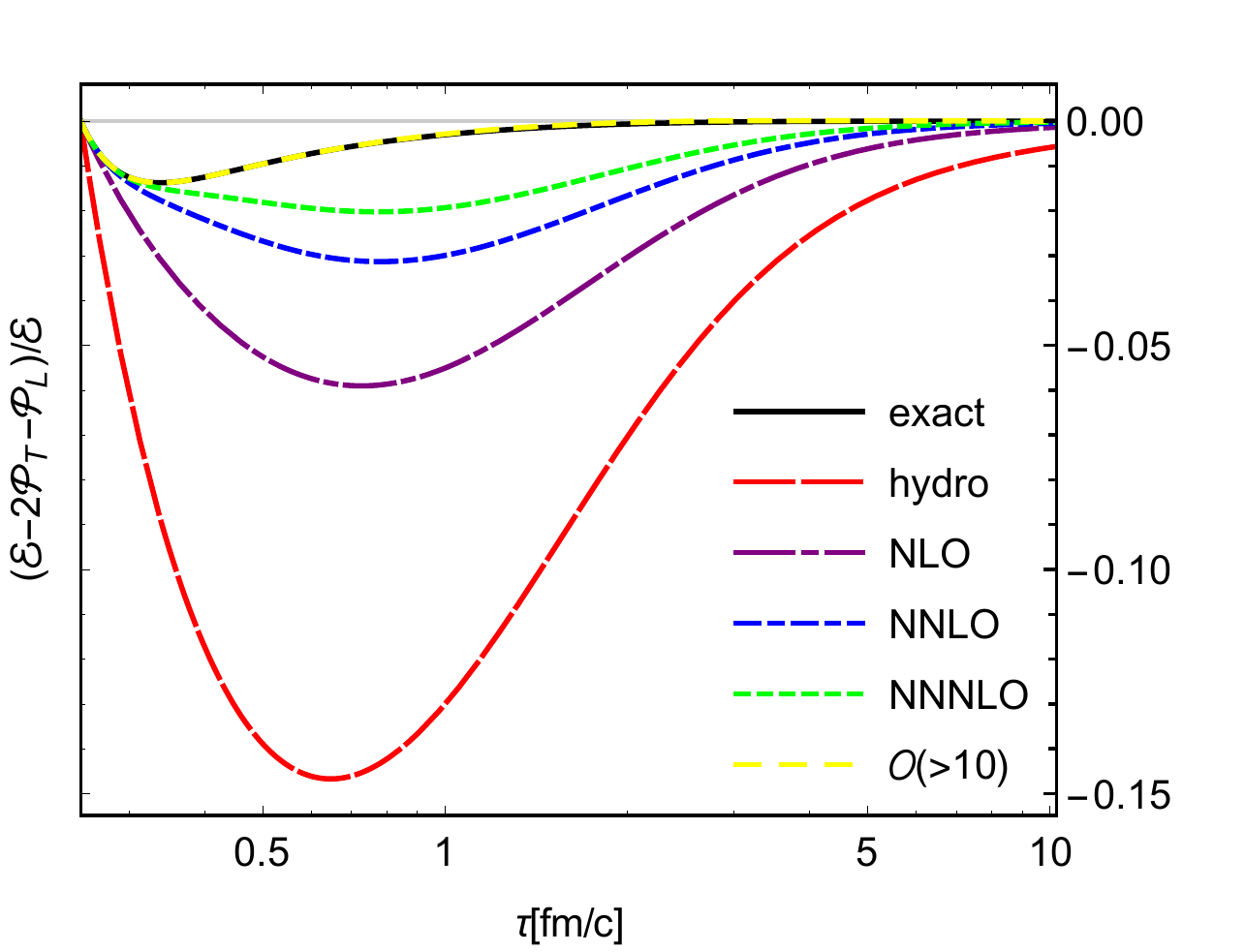}

\caption{(Color online) Trace anomaly normalized to the energy density. Color code as in Figs.~\ref{F1}. 
}
\label{F3}
\end{figure}
%


\noindent \textsl{4.Conclusions.} It is not necessary to assume a gradient expansion to derive the equation of second-order viscous hydrodynamics. It is possible to generalize the method of moments using the Wigner distribution (the quantum precursor of the distribution function) ad its evolution instead of the relativistic Boltzmann equation. The expansion must be modified to systematically avoid the infrared divergences which would otherwise appear at all orders, even for states arbitrarily close to the kinetic limit. The resulting regularized hydrodynamic expansion shows a fast convergence to the exact results, and the truncation scheme doesn't depend on the value of the gradients or the deviations from ideal hydrodynamics.
\section*{Acknowledgements} I would like to thank U. Heinz and J. Noronha for the clarifying discussions and insightful comments.This work has been funded by the Deutsche Forschungsgemeinschaft (DFG, German Research Foundation) through the CRC-TR 211 ’Strong-interaction matter under extreme conditions’– project number 315477589 – TRR 211. 


\bibliography{Wigner_Hydro_1.1}

\end{document}